\documentclass[journal]{IEEEtran}

\usepackage[backend=bibtex,style=ieee,minbibnames=1,maxbibnames=4,mincitenames=1,maxcitenames=2]{biblatex}
\usepackage[table,xcdraw]{xcolor}
\usepackage{float}
\usepackage{array}
\usepackage{booktabs}
\usepackage{arydshln}
\usepackage{multirow}
\usepackage{makecell}
\usepackage{tikz}
\usepackage{pgfplots}
\usepackage{graphicx}
\usepackage{adjustbox}
\usepackage{siunitx}
\usepackage[psamsfonts]{amssymb}
\usepackage{amsmath}
\usepackage{mathtools}
\usepackage{mathrsfs}
\usepackage{amsxtra}
\usepackage{stmaryrd}
\usepackage{accents}
\usepackage{bm}
\usepackage{upgreek}
\usepackage{etoolbox}
\usepackage[shortcuts]{extdash}
\usepackage{scalerel}
\usepackage{microtype}
\usepackage[keeplastbox]{flushend}
\usepackage{soul}
\usepackage{color}
\usepackage{listings}
\usepackage[scaled=0.8]{beramono} 
\usepackage[T1]{fontenc}
\usepackage{xurl}
\usepackage[breaklinks=true,hidelinks]{hyperref}
\hypersetup{
    colorlinks   = true,
    urlcolor     = blue,
    linkcolor    = blue,
    citecolor    = blue
}
\usepackage{pdfrender}

\newcolumntype{C}[1]{>{\centering\let\newline\\\arraybackslash\hspace{0pt}}m{#1}}
 
\newcommand*{\boldcheckmark}{%
    \textpdfrender{
        TextRenderingMode=FillStroke,
        LineWidth=.5pt, 
    }{\checkmark}%
}
\definecolor{codegray}{gray}{0.9}
\begingroup\lccode`\|=`\\
\lowercase{\endgroup\def\removebs#1{\if#1|\else#1\fi}}
\newcommand{\detokenizebackslash}[1]{\expandafter\removebs\string#1}
\makeatletter
\def\@code {\futurelet\@nexttoken\@codeaux}
\def\@codeaux {\ifx\@nexttoken\egroup\else
    \ifx\@nexttoken\bgroup
    \expandafter\expandafter\expandafter\@codea\else
    \expandafter\expandafter\expandafter\@codeb\fi\fi}
\def\@codea #1{{#1}\@code}
\def\@codeb #1{\ifcat\@nexttoken a\penalty\hyphenpenalty \codehook
    #1\else \codehook{#1}\linebreak[2]\fi\@code}
\newcommand{\highlighttt}[1]{{\fontfamily{txtt}\selectfont
        \@highlighttt #1.}}
\newcommand\code{\bgroup\ttfamily\selectfont
    \afterassignment\@code\let\next= }
\makeatother
\newcommand{\myhighlightmethod}[1]{\fboxsep0pt\colorbox{codegray}{\strut#1}}
\newcommand{\codehook}[1]{\myhighlightmethod{\detokenizebackslash{#1}}}
\lstnewenvironment{codeblock}{%
    \lstset{backgroundcolor=\color{codegray},
        frame=single,
        framerule=0pt,
        basicstyle=\ttfamily,
        mathescape=true,
        linewidth=0.49\textwidth,
        columns=fullflexible}}{}
\DeclareFontFamily{U}{mathx}{\hyphenchar\font45}
\DeclareFontShape{U}{mathx}{m}{n}{
	<5> <6> <7> <8> <9> <10>
	<10.95> <12> <14.4> <17.28> <20.74> <24.88>
	mathx10
}{}
\DeclareSymbolFont{mathx}{U}{mathx}{m}{n}
\DeclareFontSubstitution{U}{mathx}{m}{n}
\DeclareMathAccent{\widecheck}{0}{mathx}{"71}
\DeclareMathAccent{\wideparen}{0}{mathx}{"75}
\makeatletter
\let\save@mathaccent\mathaccent
\newcommand*\if@single[3]{%
\setbox0\hbox{${\mathaccent"0362{#1}}^H$}%
\setbox2\hbox{${\mathaccent"0362{\kern0pt#1}}^H$}%
\ifdim\ht0=\ht2 #3\else #2\fi
}
\newcommand*\rel@kern[1]{\kern#1\dimexpr\macc@kerna}
\newcommand*\widebar[1]{\@ifnextchar^{{\wide@bar{#1}{0}}}{\wide@bar{#1}{1}}}
\newcommand*\wide@bar[2]{\if@single{#1}{\wide@bar@{#1}{#2}{1}}{\wide@bar@{#1}{#2}{2}}}
\newcommand*\wide@bar@[3]{%
\begingroup
\def\mathaccent##1##2{%
\let\mathaccent\save@mathaccent
\if#32 \let\macc@nucleus\first@char \fi
\setbox\z@\hbox{$\macc@style{\macc@nucleus}_{}$}%
\setbox\tw@\hbox{$\macc@style{\macc@nucleus}{}_{}$}%
\dimen@\wd\tw@
\advance\dimen@-\wd\z@
\divide\dimen@ 3
\@tempdima\wd\tw@
\advance\@tempdima-\scriptspace
\divide\@tempdima 10
\advance\dimen@-\@tempdima
\ifdim\dimen@>\z@ \dimen@0pt\fi
\rel@kern{0.6}\kern-\dimen@
\if#31
	\overline{\rel@kern{-0.6}\kern\dimen@\macc@nucleus\rel@kern{0.4}\kern\dimen@}%
	\advance\dimen@0.4\dimexpr\macc@kerna
	\let\final@kern#2%
	\ifdim\dimen@<\z@ \let\final@kern1\fi
	\if\final@kern1 \kern-\dimen@\fi
\else
	\overline{\rel@kern{-0.6}\kern\dimen@#1}%
\fi
}%
\macc@depth\@ne
\let\math@bgroup\@empty \let\math@egroup\macc@set@skewchar
\mathsurround\z@ \frozen@everymath{\mathgroup\macc@group\relax}%
\macc@set@skewchar\relax
\let\mathaccentV\macc@nested@a
\if#31
	\macc@nested@a\relax111{#1}%
\else
	\def\gobble@till@marker##1\endmarker{}%
	\futurelet\first@char\gobble@till@marker#1\endmarker
	\ifcat\noexpand\first@char A\else
		\def\first@char{}%
	\fi
	\macc@nested@a\relax111{\first@char}%
\fi
\endgroup
}
\makeatother

\newcommand{\norm}[1]{\big\lVert{#1}\big\rVert}

\newcommand{\DOWNLOADURL}{%
    https://github.com/facebookresearch/EasyComDataset%
}

\begin{document}

\title{\huge EasyCom: An Augmented Reality Dataset to Support Algorithms for Easy Communication in Noisy Environments}

\author{%
 	\url{\DOWNLOADURL}\vskip1.0em%
	\IEEEauthorblockN{%
		Jacob~Donley\IEEEauthorrefmark{1},
		Vladimir~Tourbabin\IEEEauthorrefmark{1},
		Jung-Suk~Lee\IEEEauthorrefmark{1},
		Mark~Broyles\IEEEauthorrefmark{1},
	  \\Hao~Jiang\IEEEauthorrefmark{1},
		Jie~Shen\IEEEauthorrefmark{2},
		Maja~Pantic\IEEEauthorrefmark{2},
		Vamsi~Krishna~Ithapu\IEEEauthorrefmark{1},
		Ravish~Mehra\IEEEauthorrefmark{1}%
	}%
	\vskip0.5em%
	\IEEEauthorblockA{%
		\IEEEauthorrefmark{1}%
		Facebook Reality Labs Research, USA.}%
	\qquad%
	\IEEEauthorblockA{%
		\IEEEauthorrefmark{2}%
		Facebook AI Applied Research, UK.}%
	\thanks{Please send correspondence to: \href{mailto:EasyComDataset@fb.com}{EasyComDataset@fb.com}}%
}

\maketitle

\begin{abstract}
	
    Augmented Reality (AR) as a platform has the potential to facilitate the reduction of the \emph{cocktail party effect}.
    Future AR headsets could potentially leverage information from an array of sensors spanning many different modalities.
    Training and testing signal processing and machine learning algorithms on tasks such as beam-forming and speech enhancement require high quality representative data.
    To the best of the author's knowledge, as of publication there are no available datasets that contain synchronized egocentric multi-channel audio and video with dynamic movement and conversations in a noisy environment.
    In this work, we describe, evaluate and release a dataset that contains over 5 hours of multi-modal data useful for training and testing algorithms for the application of improving conversations for an AR glasses wearer.
    We provide speech intelligibility, quality and signal-to-noise ratio improvement results for a baseline method and show improvements across all tested metrics.
    The dataset we are releasing contains AR glasses egocentric multi-channel microphone array audio, wide field-of-view RGB video, speech source pose, headset microphone audio, annotated voice activity, speech transcriptions, head bounding boxes, target of speech and source identification labels.
    We have created and are releasing this dataset to facilitate research in multi-modal AR solutions to the cocktail party problem.
    
\end{abstract}



\IEEEpeerreviewmaketitle

\section{Introduction} \label{sec:Introduction}

\IEEEPARstart{T}{he} ability to communicate in noisy environments is a challenge for many people and has been a popular research topic for many decades.
The difficulty that arises is often referred to as the \emph{cocktail party effect}~\cite{cherry_experiments_1953,bronkhorst_cocktail_2000}.
As our world moves towards a new era of computing in augmented reality (AR) and virtual reality (VR) there are many benefits to be had from an all day wearable computing device, including the potential to eliminate the cocktail party effect once and for all.
The biggest obstacle to developing systems to overcome this effect is a lack of realistic data that reflects the specific scenario and provides multi-modal information.
Machine learning and signal processing algorithms can benefit from leveraging high quality multi-sensor data,
however, until now, there has not been a single dataset that provides all relevant sensor data from an AR headset in an environment that induces the cocktail party effect.

A critically differentiating aspect of an AR headset-based dataset is its egocentric nature and the associated dynamic movement.
There are few egocentric datasets publicly available that also provide data useful for solving the cocktail party effect.
The Epic Kitchens dataset~\cite{Damen2018EPICKITCHENS,damen2020rescaling}, originally released in 2018 and an extended version released in 2020, contains egocentric video and single channel audio recordings of activities that took place in kitchens.
The environments that were recorded do not contain noise or conversations that are useful for solving the cocktail party problem.
The COSINE speech dataset~\cite{stupakov2009cosine}, released in 2009, contains egocentric audio recordings with seven microphones and in naturally noisy environments.
However, the dataset does not contain other sensor modalities, such as video, nor does it contain annotated labels or data from a head mounted device.
The EgoCom dataset~\cite{northcutt2020egocom}, released in 2020, contains egocentric video and two channel audio recordings of conversations from pairs of glasses.
However, the dataset does not contain acoustic noise necessary for the problem, there is no clock synchronization between devices and the multi-channel microphone data is lossy compressed.
The Amazon Dinner Party Corpus (DiPCo)~\cite{vansegbroeck2019dipco}, released in 2019, contains 39 channels of audio as well as annotations to specifically help address the cocktail party problem in the form of a dinner party.
The CHiME-5~\cite{barker2018fifth} and CHiME-6~\cite{watanabe2020chime6} challenge datasets, released in 2018 and re-run in 2019, respectively, contain 32 audio channels of 4 participants having a dinner party in private homes.
The DiPCo, CHiME-5 and CHiME-6 datasets, however, do not contain any video or rigid wearable microphone array recordings.
The lack of egocentric perspective limits the use of the datasets and does not facilitate solving the cocktail party problem for users of wearable devices.

\setlength{\dashlinedash}{0.2pt}
\setlength{\dashlinegap}{1.5pt}
\setlength{\arrayrulewidth}{0.2pt}

\begin{table*}[htb!]
    \begin{center}
        \caption{Types of data included in existing datasets and EasyCom.} \label{tbl:dataset_comparison_datatypes}
        \begin{tabular}
            {@{}  l  c c c c c c c c   @{}}        
            \toprule
            \multirow{2}{*}{Name}                       &
            \multirow{2}{*}{Year}                       &
            \multirow{2}{*}{\makecell{\# Mics\\(sync)}} &
            \multirow{2}{*}{Video}                      &
            \multicolumn{2}{c}{Egocentric}              &
            \multirow{2}{*}{Pose}                       &
            \multirow{2}{*}{Transcribed}                &
            \multirow{2}{*}{\makecell{Noisy\\Speech}}  \\
            & & & & Mics & Camera & & &                \\
            \midrule
            COSINE \cite{stupakov2009cosine}                                  &    2009    &    7       &               &    \checkmark &               &               &    \checkmark &    \checkmark \\ \hdashline
            CHiME-5\&6 \cite{barker2018fifth,watanabe2020chime6}              &    2018/19 &    32      &               &    \checkmark &               &               &    \checkmark &    \checkmark \\ \hdashline
            DiPCo \cite{vansegbroeck2019dipco}                                &    2019    &    39      &               &               &               &               &    \checkmark &    \checkmark \\ \hdashline
            Epic Kitchens \cite{Damen2018EPICKITCHENS,damen2020rescaling}     &    2018/20 &    1       &    \checkmark &               &    \checkmark &               &               &               \\ \hdashline
            EgoCom \cite{northcutt2020egocom}                                 &    2020    &    2       &    \checkmark &    \checkmark &    \checkmark &               &    \checkmark &               \\ \hdashline
            \bf{EasyCom}                                                      &\bf{2021}   &\bf{8 to 10}&\boldcheckmark &\boldcheckmark &\boldcheckmark &\boldcheckmark &\boldcheckmark &\boldcheckmark \\
            \bottomrule
        \end{tabular}
    \end{center}
\end{table*}

In order to reduce the cocktail party effect, it is desirable to address the problem from many different domains and types of systems, such as voice/speech activity detectors, speaker diarization algorithms, direction of arrival estimators, multi-channel beamforming algorithms, single-channel audio and audio-visual speech enhancement algorithms,  automatic speech recognition algorithms, and more.
To train and test algorithms and machine learning models for these systems, it is beneficial to use high-quality realistic multi-sensor data that has been recorded in environments that closely match that of where the cocktail party effect is present.
Given the limitation of the datasets mentioned earlier, we have collected and are releasing a dataset that contains egocentric multi-channel microphone audio, video, positional data, voice activity, speech transcriptions and face bounding boxes.
The data was collected in a noisy restaurant-like room with loudspeakers placed around the room.
Participants sitting around a table attempt to have natural conversations with each other.
The video data can be used for training and testing face detection and recognition algorithms, pose estimation techniques, lip reading, conversational dynamics predictors and more.
Sensor fusion of the audio and video modalities can be performed to solve complex problems more efficiently, such as audio visual speech enhancement.
The multi-channel audio and annotations provide high quality ground truth data and labels for intrusive metrics used as loss/cost functions or for performance analysis.
We illustrate the differences between existing datasets and the dataset released in this work in table~\ref{tbl:dataset_comparison_datatypes}, which shows the key features that exist in each of them, and we summarize the types of problems that each dataset can investigate in table~\ref{tbl:dataset_comparison_problems}.

To benchmark different systems using the dataset for the application of mitigating the cocktail party effect, we propose the following task;
enhance a target speech source by improving the signal to noise ratio (SNR), speech quality and speech intelligibility whilst not distorting the original speech.
Algorithms running on this dataset would also need to cope with the dynamic movement of the sensors that are worn by the participants and should run in real-time, or at least be causal.

As a baseline starting point, we show that by building a multi-channel acoustic beamformer that is optimized to isotropic/diffuse noise we can obtain good improvements across all three benchmarking measures with minimal distortions in the target speech.
The beamformer dynamically steers to the desired speech source of interest, is used on all different wearers of the AR glasses in the dataset and can be run in real-time.

In this paper, we first describe an example task for which we use the dataset in section \ref{sec:Task}, which we follow with a
general description of the dataset in section \ref{sec:Dataset} and where it can be obtained in section \ref{sec:Availability}.
The baseline method used to for the task is explained in section \ref{sec:BaselineMethod}.
Finally, we present and discuss the results for the baseline method in section \ref{sec:Results_and_Discussion} and conclude the paper in \ref{sec:Conclusions}.

\begin{table*}[htb!]
    \begin{center}
        \caption{Types of problems that can be investigated using existing datasets and EasyCom.} \label{tbl:dataset_comparison_problems}
        \begin{tabular}
            {@{}  l  C{1.5cm} C{1.8cm} C{1.7cm} C{1.5cm} C{1.5cm} C{1.5cm} C{1.5cm}   @{}}        
            \toprule
            Use Case                                                                    & 
            \makecell{COSINE \\ \cite{stupakov2009cosine}}                              & 
            \makecell{Epic Kitchens \\ \cite{Damen2018EPICKITCHENS,damen2020rescaling}} &
            \makecell{CHiME-5\&6 \\ \cite{barker2018fifth,watanabe2020chime6}}          & 
            \makecell{DiPCo \\ \cite{vansegbroeck2019dipco}}                            & 
            \makecell{EgoCom \\ \cite{northcutt2020egocom}}                             & 
            \bf{EasyCom} \\ 
            \midrule
            Voice/Speech Activity Detection       & \checkmark &                          & \checkmark & \checkmark & \checkmark & \boldcheckmark \\ \hdashline
            Speaker Diarization                   & \checkmark &                          & \checkmark & \checkmark & \checkmark & \boldcheckmark \\ \hdashline
            Automatic Speech Recognition (ASR)    & \checkmark &                          & \checkmark & \checkmark & \checkmark & \boldcheckmark \\ \hdashline
            Noise Robust ASR                      & \checkmark &                          & \checkmark & \checkmark &            & \boldcheckmark \\ \hdashline
            Single-Ch Speech Enhancement          & \checkmark &                          & \checkmark & \checkmark &            & \boldcheckmark \\ \hdashline
            Single-Ch Speech Separation           & \checkmark &                          & \checkmark & \checkmark & \checkmark & \boldcheckmark \\ \hdashline
            Single-Ch Noise Reduction             & \checkmark & \checkmark               & \checkmark & \checkmark &            & \boldcheckmark \\ \hdashline
            Multi-Ch Speech Enhancement           & \checkmark &                          & \checkmark & \checkmark &            & \boldcheckmark \\ \hdashline
            Multi-Ch Speech Separation            & \checkmark &                          & \checkmark & \checkmark & \checkmark & \boldcheckmark \\ \hdashline
            Multi-Ch Noise Reduction              & \checkmark &                          & \checkmark & \checkmark &            & \boldcheckmark \\ \hdashline
            Acoustic Beamforming                  & \checkmark &                          & \checkmark & \checkmark & \checkmark & \boldcheckmark \\ \hdashline
            Audio-Visual Speech Enhancement       &            &                          &            &            &            & \boldcheckmark \\ \hdashline
            Audio-Visual Speech Separation        &            &                          &            &            & \checkmark & \boldcheckmark \\ \hdashline
            Audio-Visual Noise Reduction          &            & \checkmark               &            &            &            & \boldcheckmark \\ \hdashline
            Silent Video Speech Reconstruction    &            &                          &            &            & \checkmark & \boldcheckmark \\ \hdashline
            Moving-Array Acoustic Beamforming     & \checkmark &                          & \checkmark &            & \checkmark & \boldcheckmark \\ \hdashline
            Direction of Arrival Estimation       &            &                          &            &            &            & \boldcheckmark \\ \hdashline
            Sound Source Localization and Tracking&            &                          &            &            &            & \boldcheckmark \\ \hdashline
            Acoustic SLAM                         &            &                          &            &            &            & \boldcheckmark \\ \hdashline
            Visual Face Tracking                  &            &                          &            &            & \checkmark & \boldcheckmark \\ \hdashline
            Visual Body Tracking                  &            & \checkmark               &            &            & \checkmark & \boldcheckmark \\ \hdashline
            Audio-Visual Sound Source Tracking    &            & \checkmark               &            &            & \checkmark & \boldcheckmark \\ \hdashline
            Conversational Dynamics               & \checkmark &                          & \checkmark & \checkmark & \checkmark & \boldcheckmark \\ 
            \bottomrule
        \end{tabular}
    \end{center}
\end{table*}


\section{Conversational Focus Task} \label{sec:Task}
The main goal to be accomplished by using this dataset is that of conversational focus.
The aim is for an algorithm to operate on unseen data and enhance the speech of the person who is intended to be heard by the AR glasses wearer.
The enhancement of speech is defined such that there is no noise or interference and that the target speech is undistorted from that of the close microphone signals, or at least, perceptually undistorted.
Competing speech should be suppressed to either an inaudible level or a level that is not distracting to the intended conversation.
It is also an aim to simultaneously remove all noise.
Temporal and spectral distortions or modulations should not be apparent in the enhanced speech and any switching between target speech sources should be unnoticeable.

In order for any solution to this task be practical, it is imperative that the system run causally, in real-time and with minimal latency.
When considering the physical playback mechanism, for an AR application, the levels and delay between the sound originating from the scene and from the playback device should be perceptually justified.

This task is defined and driven by the motivation to improve conversations in noisy environments, such as restaurants, bars, social gatherings, etc.
The cocktail party problem that motivates this task is a problem that occurs for both normal hearing and hard of hearing.
It is, therefore, also the goal to derive a solution that is equally performant for people at all levels of hearing ability.

\section{The EasyCom Dataset} \label{sec:Dataset}
In this section, we describe the dataset's general setting, contents and features.
The dataset was collected in a room approximately $6\mathrm{m}\times 7\mathrm{m}\times 3\mathrm{m}$ (width$\times$depth$\times$height) with a reverberation time to \SI{60}{\decibel} (RT60) of \SI{645}{\milli\second}.
A table was located in the center of the room and between 3 to 5 participants sat around the table for a session duration of approximately \SI{30}{\minute} that was hosted by a single individual, referred to as the `server'.
During each session 10 loudspeakers placed around the room at varying heights played uncorrelated restaurant-like noise where each channel signal was assigned a different loudspeaker each session.
The noise was played back at a sound pressure level (SPL) of approximately \SI{71}{\decibel} as measured at the positions of the seated participants.
There are 12 sessions included in the dataset totaling approximately \SI{5}{\hour} and \SI{18}{\minute} of conversational recordings.
Additionally, there are 3 extra sessions included, which contain minor errors, such as data frame drops.

\begin{figure}
\centering
\includegraphics[width=0.7\columnwidth]{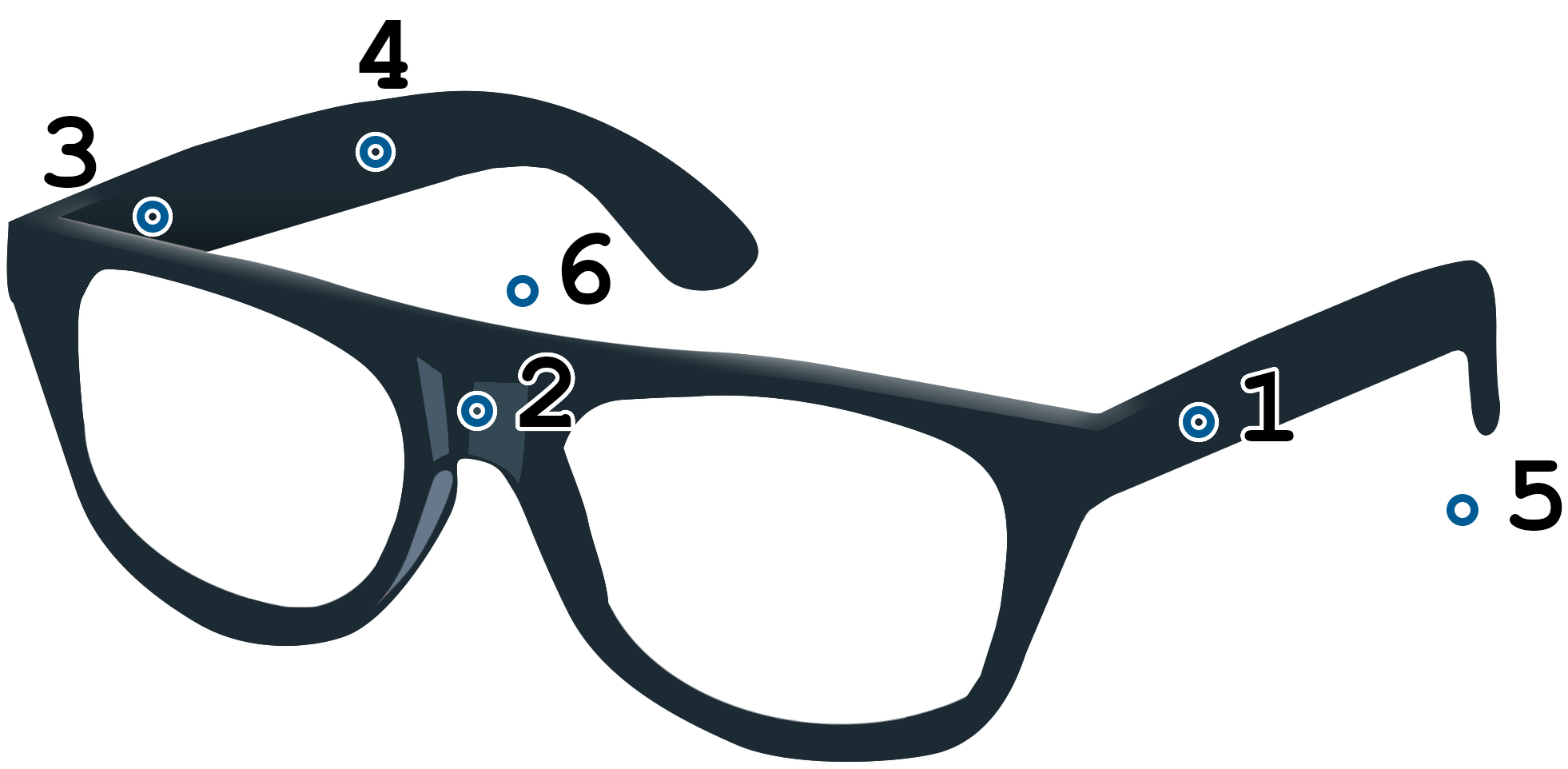}
\caption{The approximate microphone positions on the AR glasses are shown with corresponding channel number. Channels 5 and 6 are binaural microphones. This figure and the microphone positions are not a depiction of any future product.}
\label{fig:glasses_mic_positions}
\end{figure}

One participant per session was provided with an AR glasses headset that recorded multi-channel audio and video, and was tracked with tracking markers.
The multi-channel microphone array consisted of two binaural microphones (one in each ear) and an additional 4 microphones distributed around the glasses, see Fig.~\ref{fig:glasses_mic_positions} for approximate positions.
The microphones recorded audio at a sampling rate of \SI{48}{\kilo\hertz} with a bit depth of 32.
The video was recorded with a wide-angle camera at a resolution of $1920 \times 1080$ and frame rate of 20 frames per second.
The camera lens provided an approximate horizontal field of view (FOV) of $120^\circ$, vertical FOV of $66^\circ$ and diagonal FOV of $139^\circ$.
All other participants in the session, except the `server', were each provided with faux glasses that contained tracking markers and a headset microphone.
Every participant, except the `server', had their head pose optically tracked with three dimensional positions and rotations recorded at a rate of 20 frames per second.
The headset microphones were recorded at a sampling rate of \SI{48}{\kilo\hertz} with a bit depth of 32.

\begin{figure*}
    \centering
    \includegraphics[width=\textwidth]{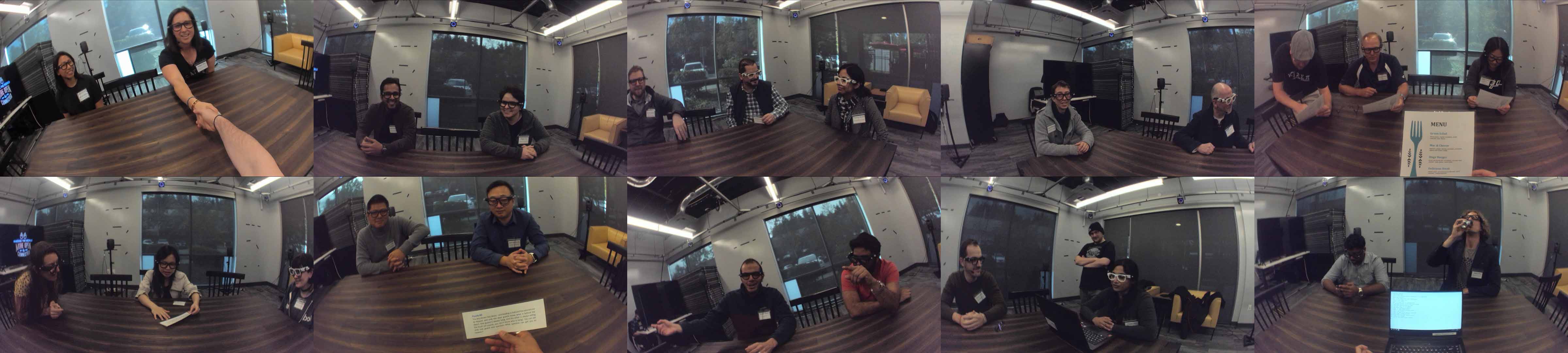}
    \caption{Example image frames from the dataset videos.}
    \label{fig:Example_Snapshot}
\end{figure*}

\subsection{Methods and Procedure}
The high level procedure undergone for recording the dataset was to initially set up the room, initialize software at the beginning of each session, prepare the hardware for participants, record participant information, start the recording process, run through several interaction tasks and stop/finish the recordings.
The initial room set up consisted of adjusting acoustic panels on walls to approximate the reverberation level of a typical occupied restaurant~\cite{rindel2012acoustical,rindel2019restaurant} and the installation of an OptiTrack\footnote{https://optitrack.com/} optical tracking system.
A table with surrounding chairs and a separate area with a central recording computer were also set up.

\begin{figure}
    \centering
    \includegraphics[width=0.8\columnwidth]{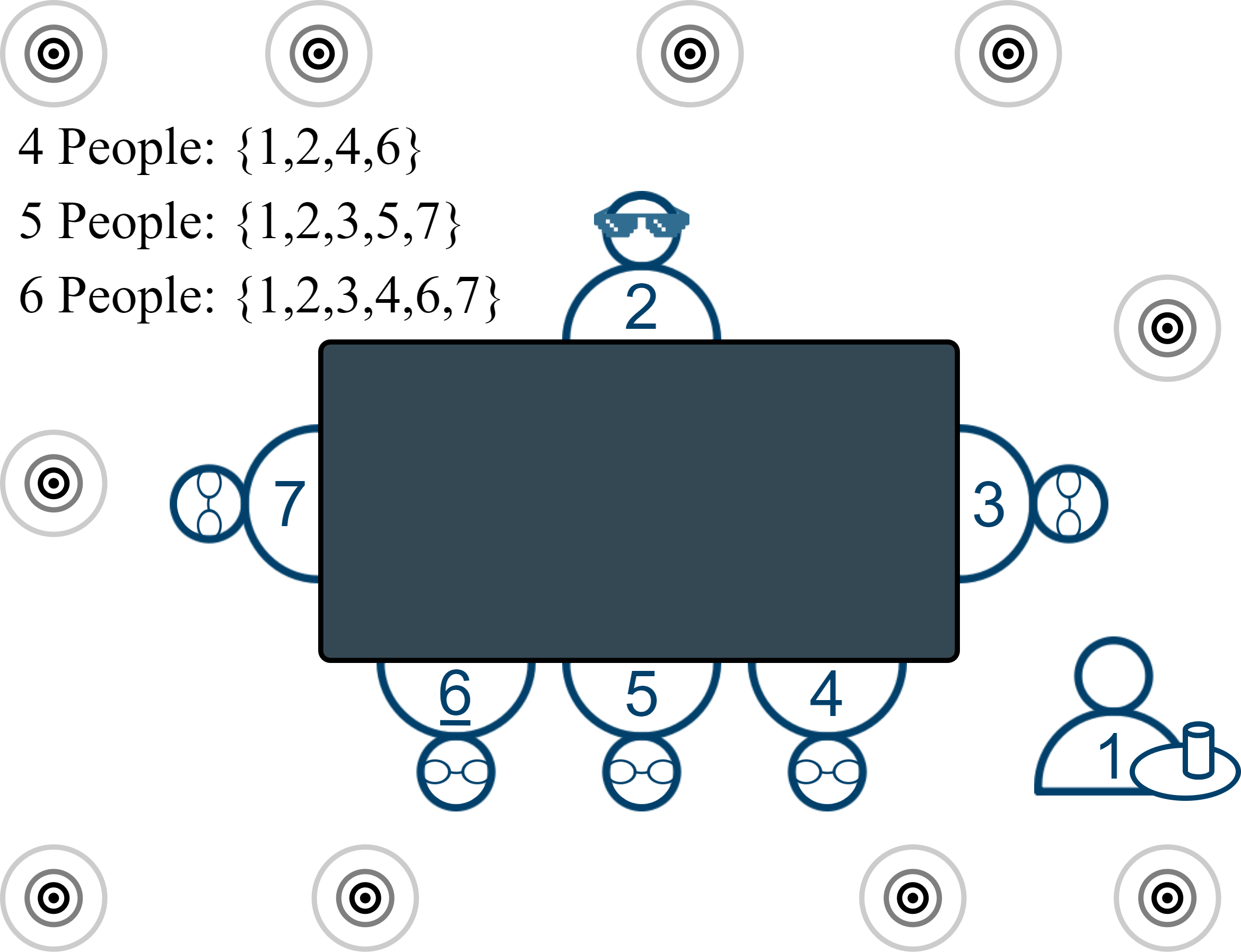}
    \caption{
        The physical layout of the recording setup is shown along with the participant ID numbers.
        The sets of IDs/locations for each given number of participants per session is also shown.
        The concentric rings indicate approximate loudspeaker locations around the room.
        The distance from participant ID 2 to ID's 3, 4, 5, 6 and 7 was approximately \SI{1.0}{\meter}, \SI{1.2}{\meter}, \SI{1.1}{\meter}, \SI{1.2}{\meter} and \SI{1.0}{\meter}, respectively.
        Dimensions are not to scale.}
    \label{fig:Dataset_Setup}
\end{figure}

Prior to recording each session the recording computer software was initialized and prepared for recording.
The participants were seated based on the total number of participants for the given session (e.g. see Fig.~\ref{fig:Dataset_Setup}).
The AR glasses hardware containing the active sensors was fitted to one of the participants.
All other participants were then fitted with a headset Rode microphone and faux AR glasses mounted only with OptiTrack infra-red reflectors (optical tracking markers).
The participant wearing the AR glasses was asked to face each other participant while an image of the participants faces were captured.
All participants were provided with faux names and occupations, and the interactions procedure was described to them.
The sound level of the restaurant-like noise was gradually increased to the specified SPL to approximate that of a typical occupied restaurant~\cite{rindel2019restaurant} and then the data recording was started immediately.

The participants began the interactions tasks when noise started, coinciding with the beginning of the recording.
The interactions tasks were completely improvised.
The first task the participants undertook was an introduction to each other using their provided faux names and occupations.
These conversations were allowed to continue anywhere up to approximately 3.5 minutes or until the `server' determined that there was a natural end in the conversation, whichever occurred first.
The `server' then proceeded to take everyone's order from a faux menu, provided to them at that time, by asking participants what they would like to eat and drink from the menu.
During pilot phases of the data collection, it was not uncommon for participants to order in a short time, so to facilitate longer interactions the `server' would judge and follow up with further questions about the menu order.
The aim was for the menu ordering to last approximately 3 minutes.
After ordering, participants menus were collected and they were all provided with a single puzzle.
The participants were encouraged to think out loud to facilitate conversation or to ask for a new puzzle if they were stuck.
The `server' would provide hints on the puzzles if participants appeared stuck.
After solving a puzzle the `server' would provide a new puzzle and this was repeated until either 10 minutes had passed solving puzzles or the participants solved them all, whichever came first.
Following the puzzle solving task, all puzzles were collected and participants were asked to play a game of ``I spy''~\cite{}, chosen to facilitate simultaneous discussion and head movement.
If any of the participants did not know the game, one of those who did was encouraged to explain it to the rest and if not a single participant knew the game, the `server' would explain the game.
The game of ``I spy'' was allowed to continue for approximately 5 minutes, after which the `server' would interrupt for the next task.
The final task of the recording procedure was to ask each participant to read a set of 20 random command-like sentences from a list (chosen for automatic speech recognition -like tasks).
Once the sentences had finished being read by each participant the recording was stopped/finished, the devices were removed from participants and they were allowed to leave.

After recording, the dataset underwent several post-processing stages which are outlined in Appendix~\ref{appx:PostProcessing}.
Calibration data for the inter- and intra- modality relationships is included in the dataset and is described in more detail in Appendix~\ref{appx:CalibrationData}.

\subsection{Annotated Labels}
We used human annotators to label Voice Activity (VA), transcribe speech, and label Target of Speech (ToS) for the dataset and used face and head detection models to generate face and head bounding boxes for each participant in the scene.
To help aid annotations, we embedded photos of each participant with a corresponding identification (ID) number (1-7) into the videos.
In order to reduce the background restaurant-like noise, and provide annotators with a clearer audio signal for labeling, we mixed the original audio captured by the AR glasses' microphones with the headset microphone audio worn by each participant.

\subsubsection{Voice Activity}
To perform annotations, we used Facebook's internal annotation platform, and utilized a video time segmentation tool for all human annotations.
The video time segmentation tool used included an audio wave form display and allowed annotators to focus to millisecond precision in order to label the times of VA.
Customizable labels and text input were utilized for speech transcription and ToS labeling.
We defined VA as any utterance by the participant.
This included any sound generated by the participant, including speech, laughing, coughing, and hesitations.
VA annotations consisted of time segments of activity and a label to designate the corresponding participant.
Annotators were instructed to watch the video and label VA for each participant sequentially by ID number, limiting their focus to labeling one participant at a time.
Rejection options were included, which annotators used to designate videos in which personally identifiable information, inappropriate language or profanity was spoken.

\subsubsection{Speech Transcriptions and Target of Speech}
For speech transcription and ToS annotations we used the VA annotation results as predefined responses, which allowed annotators to append additional labels onto the VA annotations.
Annotators were instructed to watch the video, enter the transcribed speech into a text box for each time segment of activity and designate a label corresponding to the ToS.
In addition to the participant ID numbers (labeled as numbers 1-7) we included `Group-Targeted Speech' (labeled as 0) for when the participant was addressing everyone in the scene and `Non-Targeted Speech' (labeled as 8) when the participant did not address anyone in the group (e.g. when reading sentences aloud).
In instances where more than one participant was a target of the speech, but not the whole group, we allowed annotators to designate secondary targets using extra labels.

We provided annotators with speech transcription guidelines that would result in transcriptions similar to closed-captions.
The guidelines included capitalization of proper nouns, capitalization of the first word of every sentence and used end-of-sentence punctuation.
All words and numbers were spelled out and we used coding conventions for laughter, coughing, throat clear, unintelligible speech, hesitations, and interruptions, labeled as `L', `C', `T', `U', `H' and `-', respectively.
For mispronunciations, we enclosed the utterance with asterisks (e.g. `*cabernet sauvignon*').
Upon completion of the speech transcriptions and ToS labeling annotations we completed a quality check where annotators reviewed each other's transcriptions and ToS labels to ensure accuracy and consistency.

\subsubsection{Face Bounding Boxes}
We provide per-frame face ID annotation for all videos in the dataset.
We use RetinaFace~\cite{deng2020retinaface} to detect all faces from the videos and try to match the detected faces to the ID of the glasses worn by the participants.
During the experiment, OptiTrack was used to capture the glasses' locations in the 3D space at all time.
In order to project the glasses' 3D coordinates to the 2D image plane, we calibrate the camera using OpenCV\footnote{https://opencv.org/} to estimate its intrinsics and distortion coefficients.
In each frame, we compute the pairwise Euclidean distance between the detected faces and all glasses, and then use constrained Hungarian algorithm to obtain the best match between them.
Although this process yields accurate results in most cases, manual correction is still needed because: 
1) the face detector could produce false positives; 
2) there is no OptiTrack data for the person acting as the `server'; and 
3) the distances computed on the image plane could be misleading when one face is occluding another.
However, due to the large size of the dataset, it is impractical to perform manual correction on a frame-by-frame basis.
As a remedy, we use an intersection-over-union (IoU) -based tracker~\cite{bochinski2017high} to group the detected faces into tracklets and perform the annotation on the tracklet level.
For each tracklet, we initialize its ID to be the majority voting result of the IDs automatically assigned at each frame.
We then select one representative image from each tracklet based on head pose and facial landmark detection results given by FAN~\cite{bulat2017far} and send these images for manual annotation.
Comparing to frame-level annotation, this setup reduces the total workload by a factor of 15.
Two raters were recruited to perform this manual annotation task.
In cases when their results differed, a third rater was enlisted to make the final decision.

\subsubsection{Head Bounding Boxes}
We track people's heads simultaneously in the dataset.
Our method detects the potential targets in images and then we associate
the detections through time to generate target trajectories.
However, the head detectors are not perfect.
Our tracker thus needs to deal with the missing detections and the false alarms.
For each video, head tracking gives a set of person trajectories, which which we
automatically assign IDs using the corresponding face IDs.  

The proposed method maintains the ID record of each potential trajectory
and tries to extend it to match the current head detections. 
We detect head candidates using Yolo-v3 object detection deep network~\cite{redmon2018yolov3} trained
on the Open Images dataset~\cite{kuznetsova2020open}. Matching the trajectories and the observations
can be formulated as the following optimization problem:

\begin{align*}
& \min\sum_{(i,j)}(c_{i,j}-t)x_{i,j}\\
s.t. & \sum_{i}x_{i,j}\leq1,\sum_{j}x_{i,j}\leq1\\
& x_{i,j}\mbox{ is binary }\forall i,j.
\end{align*}
Here, $x_{i,j}$ is a binary variable that indicates whether trajectory
$i$ matches head candidate $j$; if there is a match $x_{i,j}=1$,
otherwise $x_{i,j}=0$. $c_{i,j}$ is the cost of matching trajectory
$i$ with head candidate $j$. Then matching cost $c_{i,j}$ is defined
as

\[
c_{i,j}=\norm{{\bf p}_{j}-H({\bf p}_{i})}_2+\alpha d({\bf b}_{j},{\bf b}_{i}),
\]
where ${\bf p}$ is a 4-element vector whose elements are the $x$
and $y$ coordinates of the top-left corner and bottom-right corner
of the head bounding boxes. For a trajectory, its ${\bf p}$ is determined
by its last position. To compensate for the large motion of the camera
in egocentric videos, we estimate the $2\times2$ homography $H$
between the current video frame and the previous video frame using
optical flow \cite{zach2007duality}. ${\bf b}$ is a deep feature that can distinguish between head
appearances of different people. The CNN is a modified resnet-18~\cite{}
with input size of $128\times128$ and output size of $128$. It is
trained on the Voxceleb2~\cite{chung2018voxceleb2} dataset using triplet contrastive loss with
a margin of $1$. The distance $d(\cdot,\cdot)$ is $L_{2}$ distance. 

To avoid trivial all zero solutions, we introduce a threshold, $t$. There
is a potential match if the matching cost is less than the threshold
$t$. All the positive coefficients $x$ have to be zero. Otherwise,
they can be set to zero to lower the objective. The matching
for all the negative coefficient $x$ must be maximum.
The above optimization problem can be solved efficiently using the
primal-dual method~\cite{papadimitriou1998combinatorial}.

We extend the trajectory, $i$, to include the head detection, $j$, if
the corresponding $x_{i,j}=1$. The trajectory $i$'s age is increased
by $1$ and its life is restored to the maximum value, e.g. 20. If $x_{i,j}=0$
$\forall j$, and trajectory $i$'s life is greater than 0, then it
extends to a predicted position based on $M^{-1}({\bf p}_{i})$, where
$M$ is the homography estimated based on optical flow and $M^{-1}$ is
its inverse. Trajectory $i$'s age is increased by 1 and its life
is decreased by 1. We remove the trajectories who's life is less than
zero.

To deal with false alarm head detections, we only output the trajectories
whose lengths are longer than a number, e,g. 5 frames. This can effectively
remove false alarms because they are usually not stable and tend to
generate short trajectories.

To assign an ID to each trajectory, we match each head bounding box in a trajectory to a face detection that has the maximum IoU with the head detection, and record the corresponding face ID. 
A trajectory's ID is then determined by the majority vote of the matched face IDs; each head detection in the trajectory is labeled with the trajectory ID.

\begin{figure}
    \centering
    \includegraphics[width=\columnwidth]{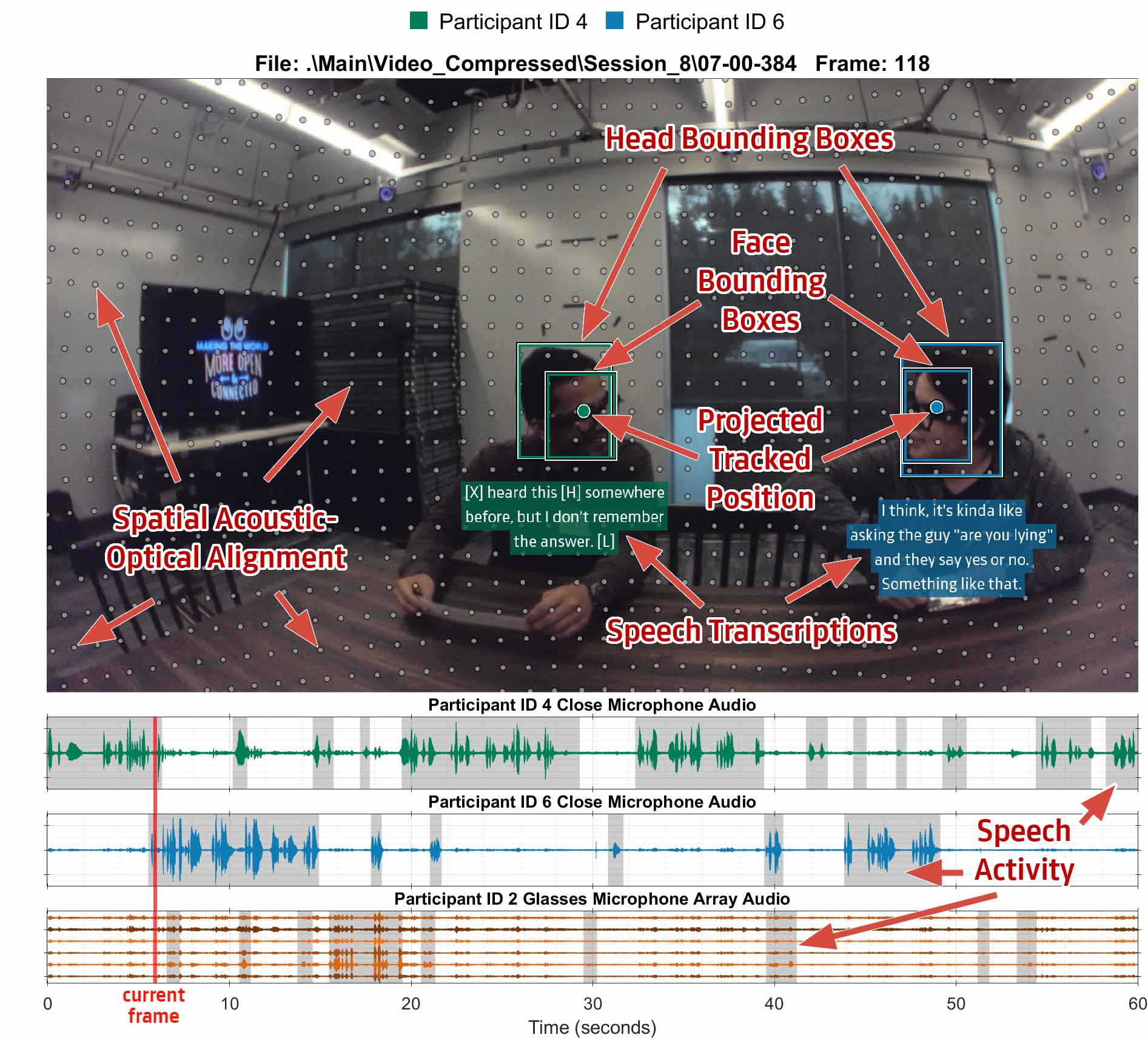}
    \caption{
        Annotations and labels are shown for an example frame in the dataset.
        The tracked poses have been projected into the image space.}
    \label{fig:annotations_and_labels}
\end{figure}

\section{Dataset Availability} \label{sec:Availability}

The dataset can be downloaded from the following address:
\begin{center}
\begin{minipage}{1.0\columnwidth}
\url{\DOWNLOADURL}
\end{minipage}
\end{center}
Please send correspondence to {\href{mailto:EasyComDataset@fb.com}{EasyComDataset@fb.com}}.
The entire dataset is available for free under the Creative Commons Attribution Non-Commercial 4.0 International public license agreement (CC-BY-NC-4.0)\footnote{\url{https://creativecommons.org/licenses/by-nc/4.0/}}.

The dataset is approximately 79\,GB in size and contains separate directories for each type of data.
The video in the dataset at the provided link is the compressed video (please contact us for details on uncompressed video).
The files are split into one minute durations for all data types.
There are 323 one minute segments totaling \SI{5}{\hour} and \SI{18}{\minute} of data.
All data and labels totals \SI{3245}{files} and approximately 1 trillion data points.
The sampling rate for each data type, and further details about the exact structure of the dataset, are included in the root directory.

\section{Baseline Method} \label{sec:BaselineMethod}

The baseline method to achieve improved conversational focus of the AR glasses wearer aims to utilize noisy signals captured by microphones of the acoustic array to generate single-channel enhanced speech of the desired participant.

Given that the enhanced speech is required to be distortionless,
we analyze a beamforming algorithm commonly referred to as the maximum directivity index (DI) beamformer as the baseline method.
The maximum DI beamformer maximizes the directvity index (and consequently the directivity factor), which is an indicator of how well diffuse noise is suppressed.
The beamformer does this while keeping the signal arriving in the target direction undistorted.
The maximum DI beamformer derivation can be formulated as an optimization problem with the following
\begin{align}
\underset{\mathbf{h}(\omega)} {\min} \quad \mathbf{h}^{\mathrm{H}}(\omega) \mathbf{R}(\omega) \mathbf{h}(\omega) \text{,}
\qquad \text{s.t. }
\mathbf{h}^{\mathrm{H}}(\omega) \mathbf{d}(\omega) = g
\end{align}
where $\omega$ denotes angular frequency and $(\cdot)^{\mathrm{H}}$ is a Hermitian transpose.
Considering an $\textit{N}$-microphone acoustic array, $\mathbf{h}(\omega)$ denotes a beamformer coefficient vector of size $\textit{N} \times 1$ and $\mathbf{d}(\omega)$ is a vector of size $\textit{N} \times 1$ that represents the acoustic transfer functions from the desired participant to the microphones on the acoustic array.
$\mathbf{R}(\omega)$  is the $\textit{N} \times \textit{N}$ multichannel covariance matrix of a spherically isotropic noise field with unit power spectral density. The solution of the optimization problem above is a coefficient vector of the maximum DI beamformer $\mathbf{h}_{\mathrm{maxDI}}(\omega)$, which can be obtained with
\begin{align}
\mathbf{h}_{\mathrm{maxDI}}(\omega) =  \frac{g^{\ast} \mathbf{R}^{-1}(\omega) \mathbf{d}(\omega)}{\mathbf{d}^H(\omega)\mathbf{R}^{-1}(\omega)\mathbf{d}(\omega)}
\end{align}
where $g$ is the constraint imposed on the beamformer output to satisfy the distortionless condition of the maximum DI beamformer and $(\cdot)^{\ast}$ indicates conjugation. If $g$ is set to be a transfer function that represents an acoustic path from the desired talker to one of the microphones on the AR glasses, then the beamformer will retrieve the speech captured by that specific microphone without distortion.

In order to compute $\mathbf{h}_{\mathrm{maxDI}}(\omega)$, two main components, $\mathbf{R}(\omega)$ and $\mathbf{d}(\omega)$, need to be consistently available while processing the microphone signals.
Theoretically $\mathbf{R}(\omega)$  can be  derived by spatial integration over a sphere, such as
\begin{align}
\mathbf{R}(\omega) = \frac{1}{4\pi} \int_{\phi} \int_{\theta} \mathbf{d}(\phi, \theta, \omega) \mathbf{d}^{\mathrm{H}}(\phi, \theta, \omega) sin(\theta)d\theta d\phi
\label{eq:R_original}
\end{align}
where $\phi$ and $\theta$ are azimuthal and inclination angles, respectively. Using the set of array transfer functions (ATFs) that are readily available from the dataset (see Appendix~\ref{appx:CalibrationData}), $\mathbf{R}(\omega)$ can be approximated with
\begin{align}
\mathbf{\tilde{R}}(\omega) = \frac{1}{N_{ATF}}\sum_{n} \sum_{m} \mathbf{d}_{ATF}(\phi_{n}, \theta_{m}, \omega)\mathbf{d}^{\mathrm{H}}_{ATF}(\phi_{n}, \theta_{m}, \omega)
\end{align}
where $N_{ATF}$ is the total number of ATFs in the set. $\phi_{n}$ and $\theta_{m}$ are the azimuthal and the inclination angles associated with a set of discrete points sampling the sphere for which an ATF vector $\mathbf{d}_{ATF}(\phi_{n}, \theta_{m}, \omega)$ is measured.

As both the desired participant and the AR glasses wearer are moving constantly relative to one another, $\mathbf{d}(\omega)$ is assumed to change over time. Hence the estimate of $\mathbf{d}(\omega)$ needs to be updated over time as opposed to $\mathbf{R}(\omega)$ whose value approximates a weakly time-varying value. We use the output of the OptiTrack tracking system as the estimated positions of the desired participant and the AR glasses
wearer. We then calculate the relative angle of the desired participant from the AR glasses wearer. Once the relative angle is estimated, $\mathbf{d}_{ATF}(\omega)$ that is closest to the relative angle is chosen from the set of ATFs and used as the estimate of $\mathbf{d}(\omega)$.

The processing of the baseline method is conducted on a frame by frame basis. At every frame, the beamformer coefficient vector, $\mathbf{h}_{\mathrm{maxDI}}(\omega)$, is updated and frames of multi-channel audio samples are filtered by $\mathbf{h}_{\mathrm{maxDI}}(\omega)$ in a weighted overlap-add (WOLA) procedure.


\section{Baseline Results and Discussion} \label{sec:Results_and_Discussion}

\begin{table*}
	\centering
	\caption{\label{tab:resultsN} Scores for the reference microphone signal and the baseline method output.}
	\begin{tabular}
		{@{} l c c c c c c c c c c c c @{}}        
		\toprule
									     & \multirow{2}{*}{Test Case}                     
									     & \multirow{2}{*}{\makecell{SNR\\\cite{brookes1997voicebox}}}
									     & \multirow{2}{*}{\makecell{SegSNR\\\cite{brookes1997voicebox}}}
									     & \multirow{2}{*}{\makecell{SDR\\\cite{vincent2006performance}}}
									     & \multirow{2}{*}{\makecell{SI-SDR\\\cite{roux2019sdr}}}
									     & \multirow{2}{*}{\makecell{STOI\\\cite{taal2011algorithm}}}
									     & \multirow{2}{*}{\makecell{ESTOI\\\cite{jensen2016algorithm}}}
									     & \multirow{2}{*}{\makecell{HASPI\\\cite{kates2021haspiv2}}}
									     & \multirow{2}{*}{\makecell{SIIB\\\cite{vankuyk2018instrumental}}}
									     & \multirow{2}{*}{\makecell{PESQ\\\cite{ITU_PESQ_2003}}}
									     & \multirow{2}{*}{\makecell{HASQI\\\cite{kates2014hasqiv2}}}
									     & \multirow{2}{*}{\makecell{ViSQOL\\\cite{chinen2020visqol}}} \\
									     \\
		\midrule
		Reference Mic                    & \multirow{2}{*}{Noise}                         &    $-9.27 $  &    $-14.2 $  &    $-8.98 $  &    $-17.5 $  &    $ 0.504$  &    $ 0.321$  &    $ 0.876$  &    $ 110  $  &    $ 1.17 $  &    $ 0.268$  &    $ 1.64 $ \\
		Baseline Method                  &                                                &$\bf{-6.62 }$ &$\bf{-10.7 }$ &$\bf{-7.79 }$ &$\bf{-14.7 }$ &$\bf{ 0.590}$ &$\bf{ 0.408}$ &$\bf{ 0.927}$ &$\bf{ 146  }$ &$\bf{ 1.27 }$ &$\bf{ 0.319}$ &$\bf{ 1.68 }$\\
        \midrule
	    Reference Mic                    & \multirow{2}{*}{\makecell{Noise +\\Interferer}}&    $-13.3 $  &    $-15.9 $  &    $-14.3 $  &    $-26.2 $  &    $ 0.462$  &    $ 0.303$  &    $ 0.720$  &    $ 107  $  &$\bf{ 1.17 }$ &    $ 0.197$  &    $ 1.65 $ \\
	    Baseline Method                  &                                                &$\bf{-10.1 }$ &$\bf{-12.2 }$ &$\bf{-12.9 }$ &$\bf{-23.4 }$ &$\bf{ 0.544}$ &$\bf{ 0.379}$ &$\bf{ 0.830}$ &$\bf{ 139  }$ &    $ 1.17 $  &$\bf{ 0.249}$ &$\bf{ 1.68 }$\\
		\bottomrule
	\end{tabular}
\end{table*}

To analyze the baseline method performance, we consider signal-to-noise ratios, speech intelligibility and speech quality.
We use 11 intrusive instrumental objective metrics as measures of the performance. 
Three metrics are related to speech quality, four metrics are related to speech intelligibility and four are related to SNR.

We use SNR as one of the metrics~\cite{brookes1997voicebox}, which is well established in the literature. We define the SNR as the ratio of the desired target source signal to all other sounds, where we consider all other sounds as undesired noise.
Another metric used is segmental SNR (SegSNR), which is similar to SNR but the mean SNR is only computed over segments where the target source signal is active~\cite{brookes1997voicebox}.
The Signal to Distortion Ratio (SDR) metric is also used and follows a similar definition to SNR but is computed using the implementation described in~\cite{vincent2006performance}.
The Scale-Invariant SDR (SI-SDR) is the last of the SNR related metrics used~\cite{roux2019sdr}. The SI-SDR is a variant of SDR that was designed to address some assumptions made in the SDR implementation from~\cite{vincent2006performance}.

Speech intelligibility was investigated using the Short-Time Objective Intelligibility (STOI) metric, which has high correlation with time-frequency weighted noisy speech and is computed on short-time segments.
A more recent version of STOI, known as Extended STOI (ESTOI) was also used in the evaluation. ESTOI has been reported to, additionally, accurately predict intelligibility when highly modulated noise is present.
The Hearing-Aid Speech Perception Index (HASPI) version 2 was another intelligibility metric used~\cite{kates2021haspiv2}. HASPI is a metric that estimates intelligibility using a model of the auditory periphery and is valid for normal-hearing and hearing-impaired listeners.
We compute HASQI assuming all participants in the dataset have normal hearing and are speaking at levels of approximately \SI{71}{\decibel}~$\mathrm{SPL}$.
The last of the intelligibility metrics used is Speech Intelligibility In Bits (SIIB)~\cite{vankuyk2018instrumental}, which estimates the amount of information shared between a talker and a listener in bits.
SIIB has been show to have higher correlation to intelligibility than STOI, ESTOI and HASPI version 1~\cite{kates2021haspiv1} across many different datasets~\cite{vankuyk2018evaluation}.

For speech quality estimation we use the Perceptual Evaluation of Speech Quality (PESQ) metric~\cite{ITU_PESQ_2003}, which is widely available and commonly used for speech quality evaluations, although it was originally designed for telephony applications.
The Hearing-Aid Speech Quality Index (HASQI) version 2 was also used to evaluate speech quality and, like HASPI, is based on a model of the auditory periphery, taking into account the effects of hearing loss. 
We compute HASQI with the same assumptions we make with HASPI. 
Finally, the last metric used in the evaluation is the Virtual Speech Quality Objective Listener (ViSQOL) version 3 metric~\cite{chinen2020visqol}, which has been shown to have higher correlation to speech quality than PESQ on several datasets~\cite{hines2015visqol}.

The metric scores are obtained on the EasyCom dataset for all possible target cases (each participant is marked as the desired participant at least once) and the average scores are shown in Table~\ref{tab:resultsN}.
`Reference Mic' and `Baseline Method' denote the case where the unprocessed reference microphone signal is used as the degraded signal for each of the metrics and the case where the baseline method output is used as the degraded signal input for each of the metrics, respectively.
Two subsets of signals are evaluated as test cases for all metrics, namely `Noise' and `Noise + Interferer'.
For the `Noise' test case, the desired participant’s portions of the signals (as determined by the VA labels of the dataset) are used as inputs for the metrics only when there is no competing talker present, and noise is always present.
For the `Noise + Interferer' test case, portions of the signal when the desired participant is active are used regardless of whether there is a competing talker present or not, and noise is always present.
We ignore cases where the participant wearing the AR glasses is actively talking.
In all evaluations, a minimally-noisy clean speech reference signal is used, which is the close microphone signals that are positioned next to the participant's mouths. 
These signals are not identical to the true clean speech component in the noisy speech signals captured by the AR glasses microphones and so the reported metric values may not be meaningful on their own.

Table~\ref{tab:resultsN} shows the results where we can see that in all test cases the baseline method increases all metrics over the `Reference Mic', except one.
The only metric that does not result in an improvement over the `Reference Mic' is PESQ and occurs for the `Noise + Interferer' test case, where the PESQ value for the `Reference Mic' and `Baseline Method' are $1.172$ and $1.168$, respectively.
This is likely due to the presence of interfering speech in the reference signal that is provided to the metric for the specific test case.
In Table~\ref{tab:resultsN} we see a \SIrange{1.2}{3.7}{\decibel} of average improvement in the SNR-based metrics.
A \SIrange{0.05}{0.11}{} improvement is made on average in most intelligibility-based metrics, except SIIB, which uses units of bits and shows \SIrange{32}{36}{bits} of improvement. We disregarded the units while making relative comparisons.
Of the improvements that are made in the quality-based metrics, we see a range of average improvement of \SIrange{0.03}{0.10}{}.

The challenging conditions of the dataset make it difficult to improve over the reference microphone signal, however, there are many other types of signals that could be leveraged to outdo the baseline method's results as reported here.
We call on the research community to improve on these results using the dataset for the tasks outlined in this paper.


\section{Conclusions} \label{sec:Conclusions}
In this work we have discussed and shown that existing datasets do not provide sufficient data for solving the cocktail party problem from a multi-modal and egocentric point of view, which could be common place in head mounted AR devices.
Existing egocentric datasets are missing either the visual modality, pose information and/or noisy environments.

We have closed the gap in the literature and the gap in available datasets by releasing a dataset with more than five hours of synchronized multi-modal egocentric noisy recordings of natural conversations.
To facilitate accelerated research in the area we have open-sourced the high quality dataset with various annotated labels in the different modalities.

We have proposed a benchmarking task for the dataset and have analyzed a baseline method to address the task.
The baseline method leverages acoustic and positional information to enhance targeted speech in real-time.
Other included modalities could also be leveraged when addressing the benchmarking task.
Our baseline method results in substantial and consistent improvements in speech quality, intelligibility and signal-to-noise ratio -based metrics whilst not distorting the target speech.
Our method can also instantaneously switch between target speech sources.

We challenge the research community to outperform our baseline using the benchmarking task on the provided dataset to help solve the cocktail party problem for AR.


\appendices

\section{Post-Processing} \label{appx:PostProcessing}

After the recording of the raw data, several post-processing steps took place to improve the quality of the dataset.
The post-process reduced the size of the dataset, aligned clean speech reference signals and facilitated more accurate annotations.

\subsection{Video Compression} \label{sec:video_compression}
The storage size of the raw uncompressed videos, at several terabytes, is substantially larger than that of the visually lossless (lossy) compressed version, at approximately 42 gigabytes.
The video compression was performed using FFMPEG~\cite{}.
The videos were encoded from uncompressed MPEG in an AVI container to an MPEG4 file format.
The MPEG4 video was compressed using the x264 library with settings using the `veryslow' preset and a Constant Rate Factor (CRF) of 17.

The adaptive B-frame decision method was optimal with 8 B-frames between I and P.
The direction motion vector prediction was automatic and integer pixel motion estimation method was an uneven multi-hexagon search.
The maximum motion vector search range was 24.
All partitions of the P, B and I -macroblocks were considered.
The number of reference frames was 16.
The sub-pixel motion estimation and mode decision was with quantization parameter and trellis RD quantization, enabled on all mode decisions.
There were 60 frames for the frame-type lookahead.

Two channels of the recorded audio were embedded in the compressed video.
The binaural microphone audio from channel five (left) and channel six (right) of the microphone array was embedded.

\subsection{Audio Alignment} \label{sec:audio_alignment}
The AR glasses microphone array and headset microphones were sampled with two different hardware clocks.
Each frame of \SI{2400}{samples} of audio was timestamped using a central clock.
The headset microphone recordings were first coarsely aligned with the time stamps.
The alignment was then further refined using the absolute peak of a generalized cross-correlation with phase transform (GCC-PHAT)~\cite{1162830} using the headset microphone signals and a reference microphone on the AR glasses.

\section{Calibration Data} \label{appx:CalibrationData}
To facilitate effective use of the data across different modalities, calibration data is provided.
The calibration data allows for efficient use of the acoustic array, linking the acoustic and visual modalities, reducing optical distortions and refining tracking estimates.
None of the provided calibration data that is described in the following is a depiction of any future product.

\subsection{Acoustic Array Transfer Functions}
In order to effectively compute spatial enhancement filters or beamforming filters, the transfer function from each microphone to a point in far-field space is required.
This set of transfer functions is commonly called the steering vector, array manifold or array transfer functions (ATFs), the latter is the term used in this work.
The ATFs included in this dataset were measured on a head and torso simulator in an anechoic chamber for a discrete set of positions on a sphere.

\subsection{Spatial Acoustic-Optical Alignment}
The acoustic array has a spatial response and, hence, the ability to localize sound sources in space.
The camera spatially samples photons from the surrounding space and can be correlated with the acoustic space if both spaces are aligned.
In order to align these two modalities, the AR glasses were placed on a head and torso acoustic simulator in a specialized acoustic setup.
The discrete acoustic source locations in azimuth and elevation were marked in the camera field of view and labeled as the corresponding image pixel coordinates.

\subsection{Optical Camera Intrinsics}
The camera lens path introduces typical lens distortions, which change the way the flat image sensor maps to world coordinates.
To compensate for this lens warping, the camera intrinsics and lens de-warp filter parameters were measured.
Images of a standard checkerboard were taken with the camera and then used to produce the de-warping filters.

\subsection{Tracking Marker Locations}
The OptiTrack tracking system used multiple passive infra-red reflective markers to track the participants in the scene.
Multiple tracking markers were used per pair of glasses, five for the AR glasses recording the scene and four for the other glasses.
The resulting tracked position included in the dataset is the center of mass of the tracked marker locations.
In order to re-align the tracked positions to a specific marker, the relative positions of the markers and the center of mass are provided in the dataset.


\section*{Acknowledgments} \label{sec:Acknowledgments}

We would like to thank the research assistant for their excellent work helping collect this dataset (whose name is anonymous for privacy reasons).
We would also like to thank everyone who gave valuable feedback on the dataset.
Lastly, we extend our thanks to all participants involved in the data collection.


\renewcommand*{\bibfont}{\footnotesize}
\setlength\bibitemsep{0.2em}
\printbibliography

@article{cherry_experiments_1953,
	title = {Some Experiments on the Recognition of Speech, with One and with Two Ears},
	volume = {25},
	pages = {975--979},
	number = {5},
	journal = {J. Acoust. Soc. Am.},
	author = {Cherry, E. Colin},
	year = {1953},
	month = sep
}

@article{bronkhorst_cocktail_2000,
	title = {The cocktail party phenomenon: A review of research on speech intelligibility in multiple-talker conditions},
	volume = {86},
	shorttitle = {The cocktail party phenomenon},
	pages = {117--128},
	number = {1},
	journal = {Acustica},
	author = {Bronkhorst, Adelbert W.},
	year = {2000}
}

@inproceedings{Damen2018EPICKITCHENS,
	title={Scaling Egocentric Vision: The EPIC-KITCHENS Dataset},
	author={Damen, Dima and Doughty, Hazel and Farinella, Giovanni Maria  and Fidler, Sanja and 
	Furnari, Antonino and Kazakos, Evangelos and Moltisanti, Davide and Munro, Jonathan 
	and Perrett, Toby and Price, Will and Wray, Michael},
	booktitle={Eur. Conf. Comput. Vision (ECCV)},
	year={2018}
}

@ARTICLE{damen2020rescaling,
	title={Rescaling Egocentric Vision},
	author={Damen, Dima and Doughty, Hazel and Farinella, Giovanni Maria and Furnari, Antonino 
	and Ma, Jian and Kazakos, Evangelos and Moltisanti, Davide and Munro, Jonathan 
	and Perrett, Toby and Price, Will and Wray, Michael},
	year= {2020},
	eprint={2006.13256},
	archivePrefix={arXiv},
	primaryClass={cs.CV}
}

@inproceedings{stupakov2009cosine,
	author={A. {Stupakov} and E. {Hanusa} and J. {Bilmes} and D. {Fox}},
	booktitle={IEEE Int. Conf. on Acoust., Speech and Signal Process. ({ICASSP})},
	title={{COSINE} - A corpus of multi-party COnversational Speech In Noisy Environments},
	year={2009},
	volume={},
	number={},
	pages={4153-4156}
}

@article{northcutt2020egocom,
	author={C. {Northcutt} and S. {Zha} and S. {Lovegrove} and R. {Newcombe}},
	journal={IEEE Trans. Pattern Anal. and Mach. Intell.},
	title={{EgoCom}: A Multi-person Multi-modal Egocentric Communications Dataset},
	year={2020},
	volume={},
	number={},
	pages={1-1}
}

@misc{vansegbroeck2019dipco,
	title={{DiPCo} -- Dinner Party Corpus}, 
	author={Maarten Van Segbroeck and Ahmed Zaid and Ksenia Kutsenko and Cirenia Huerta and Tinh Nguyen and Xuewen Luo and Björn Hoffmeister and Jan Trmal and Maurizio Omologo and Roland Maas},
	year={2019},
	eprint={1909.13447},
	archivePrefix={arXiv},
	primaryClass={eess.AS}
}

@misc{barker2018fifth,
	title={The fifth {`CHiME'} Speech Separation and Recognition Challenge: Dataset, task and baselines}, 
	author={Jon Barker and Shinji Watanabe and Emmanuel Vincent and Jan Trmal},
	year={2018},
	eprint={1803.10609},
	archivePrefix={arXiv},
	primaryClass={cs.SD}
}

@misc{watanabe2020chime6,
	title={{CHiME-6} Challenge: Tackling Multispeaker Speech Recognition for Unsegmented Recordings}, 
	author={Shinji Watanabe and Michael Mandel and Jon Barker and Emmanuel Vincent and Ashish Arora and Xuankai Chang and Sanjeev Khudanpur and Vimal Manohar and Daniel Povey and Desh Raj and David Snyder and Aswin Shanmugam Subramanian and Jan Trmal and Bar Ben Yair and Christoph Boeddeker and Zhaoheng Ni and Yusuke Fujita and Shota Horiguchi and Naoyuki Kanda and Takuya Yoshioka and Neville Ryant},
	year={2020},
	eprint={2004.09249},
	archivePrefix={arXiv},
	primaryClass={cs.SD}
}

@article{rindel2012acoustical,
	title={Acoustical capacity as a means of noise control in eating establishments},
	author={Rindel, Jens Holger},
	journal={Proc. Baltic-Nordic Acoust. Meeting},
	volume={2429},
	year={2012}
}

@article{rindel2019restaurant,
	title={Restaurant acoustics--Verbal communication in eating establishments},
	author={Rindel, JH},
	journal={Acoust. Pract.},
	volume={7},
	number={1-14},
	year={2019}
}

@inproceedings{deng2020retinaface,
	title={Retinaface: Single-shot multi-level face localisation in the wild},
	author={Deng, Jiankang and Guo, Jia and Ververas, Evangelos and Kotsia, Irene and Zafeiriou, Stefanos},
	booktitle={Proc. {IEEE}/{CVF} Conf. Comput. Vision and Pattern Recognit.},
	pages={5203--5212},
	year={2020}
}

@inproceedings{bochinski2017high,
	title={High-speed tracking-by-detection without using image information},
	author={Bochinski, Erik and Eiselein, Volker and Sikora, Thomas},
	booktitle={{IEEE} Int. Conf. Adv. Video and Signal Based Surveillance ({AVSS})},
	pages={1--6},
	year={2017}
}

@inproceedings{bulat2017far,
	title={How far are we from solving the 2d \& 3d face alignment problem?(and a dataset of 230,000 3d facial landmarks)},
	author={Bulat, Adrian and Tzimiropoulos, Georgios},
	booktitle={Proc. {IEEE} Int. Conf. Comput. Vision},
	pages={1021--1030},
	year={2017}
}

@article{redmon2018yolov3,
	title={{YOLOv3}: An incremental improvement},
	author={Redmon, Joseph and Farhadi, Ali},
	year={2018},
	eprint={1804.02767},
	archivePrefix={arXiv},
	primaryClass={cs.CV}
}

@article{kuznetsova2020open,
	title={The Open Images Dataset V4},
	author={Kuznetsova, Alina and Rom, Hassan and Alldrin, Neil and Uijlings, Jasper and Krasin, Ivan and Pont-Tuset, Jordi and Kamali, Shahab and Popov, Stefan and Malloci, Matteo and Kolesnikov, Alexander and others},
	journal={Int. J. Comput. Vision},
	pages={1--26},
	year={2020},
	publisher={Springer}
}

@inproceedings{zach2007duality,
	title={A Duality Based Approach for Realtime {TV-L1} Optical Flow},
	author={Zach, Christopher and Pock, Thomas and Bischof, Horst},
	booktitle={Joint pattern recognition symposium},
	pages={214--223},
	year={2007},
	organization={Springer}
}

@article{chung2018voxceleb2,
	title={{VoxCeleb2}: Deep Speaker Recognition},
	author={Chung, Joon Son and Nagrani, Arsha and Zisserman, Andrew},
	journal={Proc. Interspeech},
	pages={1086--1090},
	year={2018}
}

@book{papadimitriou1998combinatorial,
	title={Combinatorial Optimization: Algorithms and Complexity},
	author={Papadimitriou, Christos H and Steiglitz, Kenneth},
	year={1998},
	publisher={Courier Corporation}
}

@article{brookes1997voicebox,
	title={Voicebox: Speech processing toolbox for matlab},
	author={Brookes, Mike},
	journal={Software, available from www. ee. ic. ac. uk/hp/staff/dmb/voicebox/voicebox. html},
	volume={47},
	year={1997}
}

@ARTICLE{vincent2006performance,
	author={Vincent, E. and Gribonval, R. and Fevotte, C.},
	journal={{IEEE} Trans. Audio, Speech, Lang. Process.}, 
	title={Performance measurement in blind audio source separation}, 
	year={2006},
	volume={14},
	number={4},
	pages={1462-1469}
}

@INPROCEEDINGS{roux2019sdr,
	author={Roux, Jonathan Le and Wisdom, Scott and Erdogan, Hakan and Hershey, John R.},
	booktitle={{IEEE} Int. Conf. on Acoust., Speech and Signal Process. ({ICASSP})},
	title={{SDR} – Half-baked or Well Done?}, 
	year={2019},
	volume={},
	number={},
	pages={626-630}
}

@article{taal2011algorithm,
	title={An algorithm for intelligibility prediction of time--frequency weighted noisy speech},
	author={Taal, Cees H and Hendriks, Richard C and Heusdens, Richard and Jensen, Jesper},
	journal={{IEEE} Trans. Audio, Speech, and Lang. Process.},
	volume={19},
	number={7},
	pages={2125--2136},
	year={2011}
}

@ARTICLE{jensen2016algorithm,
	author={Jensen, Jesper and Taal, Cees H.},
	journal={{IEEE}/{ACM} Trans. Audio, Speech, Lang. Process.},
	title={An Algorithm for Predicting the Intelligibility of Speech Masked by Modulated Noise Maskers}, 
	year={2016},
	volume={24},
	number={11},
	pages={2009-2022}
}

@article{kates2021haspiv2,
	title = {The Hearing-Aid Speech Perception Index ({HASPI}) Version 2},
	journal = {Speech Commun.},
	volume = {131},
	pages = {35-46},
	year = {2021},
	issn = {0167-6393},
	author = {James M. Kates and Kathryn H. Arehart}
}

@ARTICLE{vankuyk2018instrumental,
	author={Van Kuyk, Steven and Kleijn, W. Bastiaan and Hendriks, Richard C.},
	journal={{IEEE} Signal Process. Lett.}, 
	title={An Instrumental Intelligibility Metric Based on Information Theory}, 
	year={2018},
	volume={25},
	number={1},
	pages={115-119}
}

@book{ITU_PESQ_2003,
	title={Perceptual evaluation of speech quality ({PESQ})},
	publisher={Int. Telecommun. Union ({ITU}), {ITU}-T Rec. P.862},
	year={2003}
}

@article{kates2014hasqiv2,
	author={Kates, James M. and Arehart, Kathryn H.},
	journal={J. Audio Eng. Soc.},
	title={The Hearing-Aid Speech Quality Index ({HASQI}) Version 2},
	year={2014},
	volume={62},
	number={3},
	pages={99-117},
	month={march}
}

@inproceedings{chinen2020visqol,
	title={{ViSQOL} v3: An open source production ready objective speech and audio metric},
	author={Chinen, Michael and Lim, Felicia SC and Skoglund, Jan and Gureev, Nikita and O'Gorman, Feargus and Hines, Andrew},
	booktitle={Int. Conf. Quality Multimedia Experience (QoMEX)},
	pages={1--6},
	year={2020},
	organization={IEEE}
}

@article{kates2021haspiv1,
	title = {The Hearing-Aid Speech Perception Index ({HASPI})},
	journal = {Speech Commun.},
	volume = {65},
	pages = {75-93},
	year = {2014},
	issn = {0167-6393},
	author = {James M. Kates and Kathryn H. Arehart}
}

@ARTICLE{vankuyk2018evaluation,
	author={Van Kuyk, Steven and Kleijn, W. Bastiaan and Hendriks, Richard Christian},
	journal={{IEEE}/{ACM} Trans. Audio, Speech, Lang. Process.},
	title={An Evaluation of Intrusive Instrumental Intelligibility Metrics}, 
	year={2018},
	volume={26},
	number={11},
	pages={2153-2166}
}

@article{hines2015visqol,
	title={{ViSQOL}: an objective speech quality model},
	author={Hines, Andrew and Skoglund, Jan and Kokaram, Anil C and Harte, Naomi},
	journal={{EURASIP} J. Audio, Speech, Music Process.},
	volume={2015},
	number={1},
	pages={1--18},
	year={2015},
	publisher={SpringerOpen}
}

@ARTICLE{1162830,
	author={C. {Knapp} and G. {Carter}},
	journal={{IEEE} Trans. Acoust., Speech, and Signal Process.}, 
	title={The generalized correlation method for estimation of time delay}, 
	year={1976},
	volume={24},
	number={4},
	pages={320-327}
}

\end{document}